# Neutron Knockout on Beams of $^{108,106}$Sn and $^{106}$Cd[*]

G. Cerizza

*University Of Tennessee, Department of Physics and Astronomy, Knoxville, TN 37996, USA*
[†]*E-mail: gcerizza@utk.edu*
*www.phys.utk.edu*

Characterizing the nature of single-particle states outside of double shell closures is essential to a fundamental understanding of nuclear structure. This is especially true for those doubly magic nuclei that lie far from stability and where the shell closures influence nucleo-synthetic pathways. The region around $^{100}$Sn is one of the most important due to the proximity of the N=Z=50 magic numbers, the proton-drip line, and the end of the rp-process. However, owing to the low production rates, there is a lack of spectroscopic information and no firm spin-parity assignment for ground states of odd-A isotopes close to $^{100}$Sn. Neutron knockout reaction experiments on beams of $^{108,106}$Sn and $^{106}$Cd have been performed at the NSCL. By measuring gamma rays and momentum distributions from reaction residues, the spin of ground state and first excited state for $^{107,105}$Sn have been established. The results also show a degree of mixing in the ground states of the isotopes $^{108,106}$Sn between the $d_{5/2}$ and $g_{7/2}$ single particle-states and they are compared to reaction calculations. Single-, double-, and triple-neutron knockout reactions on the $^{106}$Cd beam have been observed. The spin-parity of $^{105}$Cd is already known, therefore, the measurement of the momentum distributions of the ground and first excited states of this residue is an important validation of the technique used for the light tin isotopes.

*Keywords*: Nucleosyntesis; rp Process; Nuclear Structure; Nuclear Reaction; Shell Evolution.

## 1. Introduction and Physics Case

The region close to $^{100}$Sn is rich in physics and is especially important to our understanding of nuclear structure due to the proximity of the N=Z=50 magic numbers, the N=Z line, the proton drip-line and the end of the rp-process. These nuclei are some of the best candidates for testing the validity of shell-model predictions as they have just a few nucleons outside the $^{100}$Sn closed shell. Because of experimental difficulties, experimental information in this region is very scarce. One important step in understanding shell structure in this region is to establish the energies of single-particle states. The most neutron deficient tin nucleus in which an excited state has been measured is $^{101}$Sn [1]. A prompt γ ray

---

[*] Presented at the First International African Symposium on Exotic Nuclei IASEN2013 Ithemba LABS Cape Town, South Africa, December 2-6, 2013

correlated to β-delayed protons was observed at 172 keV. Darby et al [2] measured the $^{109}$Xe→$^{105}$Te→$^{101}$Sn α-decay chain and discovered that the majority of the α decays of $^{105}$Te populated the first excited state in $^{101}$Sn, in contrast to what had been previously observed in the α decay of $^{107}$Te, where the population of the excited state in $^{103}$Sn was very small (<1%) [3]. The conclusion of this study was that the ground state of $^{101}$Sn is not 5/2$^+$ and that is a g$_{7/2}$ neutron single particle state. In support of this hypothesis shell-model calculations for the splitting of the 7/2$^+$ - 5/2$^+$ states in the neutron-deficient, odd-mass tin isotopes were presented. A $^{100}$Sn core and d$_{5/2}$, g$_{7/2}$, h$_{11/2}$ and s$_{1/2}$ model space for protons and neutrons was assumed in the first variant (V1) with residual interactions based on AV18 [4] or N3LO [5] nucleon-nucleon potentials, while the second one (V2) used a $^{88}$Sr core with valence protons in p$_{1/2}$ and g$_{9/2}$ shells with effective interactions based on the CD-Bonn potential [6]. Figure 1 compares the results of V1 and V2 with experiment. The splitting of the d$_{5/2}$ and g$_{7/2}$ at $^{101}$Sn has been set to the experimental value of 172 keV for V1. In V2 the neutron g$_{7/2}$ is 2.43 MeV above the d$_{5/2}$ level in $^{88}$Sr [7,8] and the level crossing in $^{101}$Sn is entirely due to the properties of the effective interactions. The V1 calculations were performed for both possible orderings of the levels, thus allowing for either a 7/2$^+$ or 5/2$^+$ ground state in $^{101}$Sn. However, regardless of this ordering, the ground state of the heavier tin isotopes, most notably $^{103}$Sn, is always 5/2$^+$. By forcing a 5/2$^+$ ground state in $^{101}$Sn, the calculations overestimate the location of the first excited-state in the heavier tin isotopes by 200 keV, a 7/2$^+$ ground state, on the other hand, gives excellent agreement between theory and experiment. This odd behavior from these new calculations is a result of a strong (g$_{7/2}$)$^2$ $_{J=0}$ pairing matrix element of the residual interactions which makes the (g$_{7/2}$)$^2$ neutron configurations energetically more favorable, whereas in the conventional seniority scheme the pairing TBME's are proportional to *2j+1* and their effects cancel out. Thus, in the realistic interaction picture, the low energy states in $^{103-109}$Sn are in fact dominated by neutron correlations while the states in $^{101}$Sn follow the single particle nature of this $^{100}$Sn + 1 valence neutron nucleus. The calculations following the Hjorth-Jensen[9,10] prescription predict ground states for the even-mass neutron-deficient tin isotopes that are highly mixed, with the g$_{7/2}$ configuration dominating for $^{102}$Sn and $^{104}$Sn, and the d$_{5/2}$ configuration dominating for $^{106}$Sn and $^{108}$Sn. The α decay experiment shows that the ground state of $^{101}$Sn is similar in nature to the first excited state in $^{105}$Te. While the agreement between α-decay results and theory is compelling, neither state has an independently established spin-parity assignment. In fact, previous to the measurement at the NSCL presented here, the last odd-mass tin nucleus with a firm spin-parity assignment for its ground state was $^{109}$Sn, which is 5/2$^+$[11]. The other odd-mass tin isotopes reaching out to $^{101}$Sn were assumed to have a 5/2$^+$ ground state from extending the systematics from $^{109}$Sn. The strong pairing interaction that drives the mixing of the states for A>101 tin isotopes is a non-trivial effect that can be traced to the parameterization of the high angular momentum part of the free nucleon-nucleon interaction. Measurement of the nature of the ground states of light tin isotopes is therefore crucial not only to determine the structure of these nuclei, but even more importantly, to provide the direct and fundamental link between free nucleon-nucleon forces and their influence on the properties of heavy nuclei. For the $^{106}$Cd beam, that was present in the same experiment configuration, studies of neutron knockout reactions are a solid benchmark case to validate the light tin isotope analysis technique because the spin-parity of the ground and first excited states are known.

## 2. Experiment

In 2011 at the NSCL the residues and γ rays emerging from reactions on $^{108,106}$Sn and $^{106}$Cd beams were measured[12]. A primary beam of $^{124}$Xe at 140 MeV/u on a $^9$Be target and a momentum acceptance of 3% was used to produce $^{108}$Sn and $^{106}$Sn beams with average rates of 3x10$^3$ pps and 1x10$^2$ pps, respectively. The $^{106}$Cd beam was produced parasitically within the same settings of rigidity with an average rate of 15x10$^3$ pps. The beams impinged on a 257μm thick $^9$Be target producing reaction fragments including $^{107,105}$Sn and $^{105,104,103}$Cd. In approximately two days of run per beam setting, about 9600, 2000, and 54000 events of $^{107}$Sn, $^{105}$Sn, and $^{105}$Cd were measured, respectively. The experiment used NSCL's high-efficiency γ ray-detector array CAESAR [13], which has an efficiency of 80% for 200 keV γ rays, in conjunction with the S800 spectrograph for coincident particle γ ray spectroscopy. The reaction residues were detected and identified in the S800 focal plane detector system [14] operated in dispersion-matched mode. The neutron-knockout residues from the unreacted beam passing through the target are shown in Figure 2 for the tin case. The characterization of the reaction fragments was performed via γ spectroscopy. A Maximum Likelihood (ML) fit [15] was performed to the γ spectra to construct level schemes through γ-γ coincidences. Figure 3 shows the fit to the data up to 2.5 MeV for $^{107}$Sn. The higher energy gamma spectrum is still under investigation as γ transitions from knockout neutrons from below the N=50 shell gap may be present. For events that populate the ground and first excited states the longitudinal momentum distributions were reconstructed from the dispersive position of the residue tracked through the spectrograph. The momentum distributions obtained evidence that for knockout to the ground and first excited states in $^{107}$Sn the neutrons were in the $d_{5/2}$ and $g_{7/2}$ orbitals, respectively.
Studies of the $^{106}$Sn beam and its one-neutron knockout product $^{105}$Sn are still ongoing, although early results show the same trend as $^{108}$Sn and $^{107}$Sn.

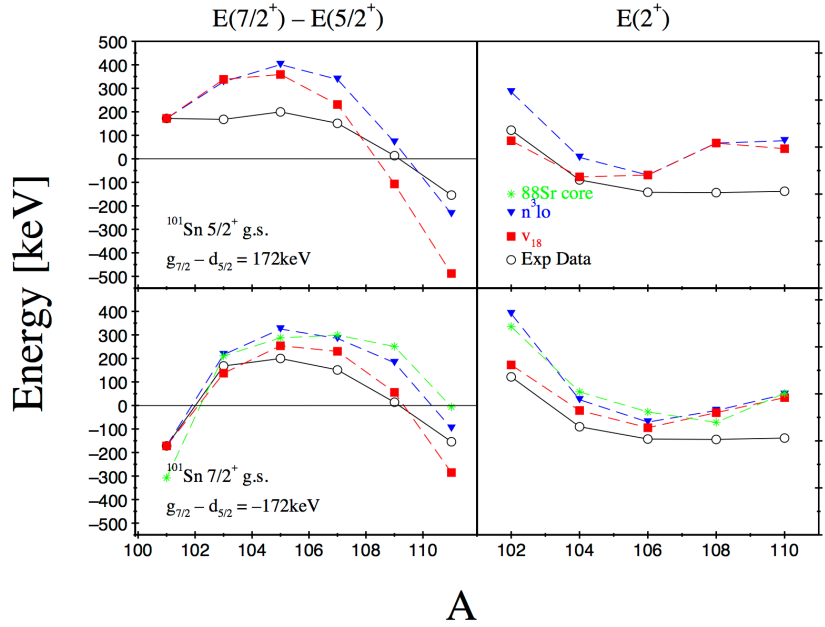

Figure 1: Results of the shell-model calculations in variants V1 and V2 for the splitting of the $7/2^+$-$5/2^+$ states in the neutron-deficient, odd-mass tin isotopes (left). In V1, the upper (lower) panel assumes the $d_{5/2}$ single-neutron level above (below) the $g_{7/2}$ level. Experimental values (circles) are taken from Darby et al. [16] and references quoted therein. On the left, E(2+) energies summary plot for even-mass tin isotopes. The two shell-model calculation variants V1 and V2 are overlaid to the data.

Preliminary inclusive cross sections of about 60 mb and 30mb have been measure for one-neutron knockout reactions of $^{108}$Sn and $^{106}$Sn, respectively.

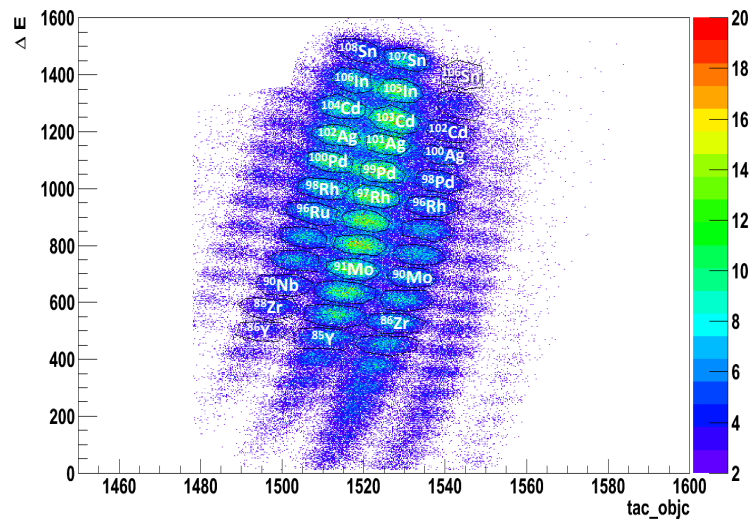

Figure 2: Summary of the reaction fragments of the $^{108}$Sn beam impinged on a $^9$Be target. The isotopes have been identified with the S800 spectrograph via γ spectroscopy.

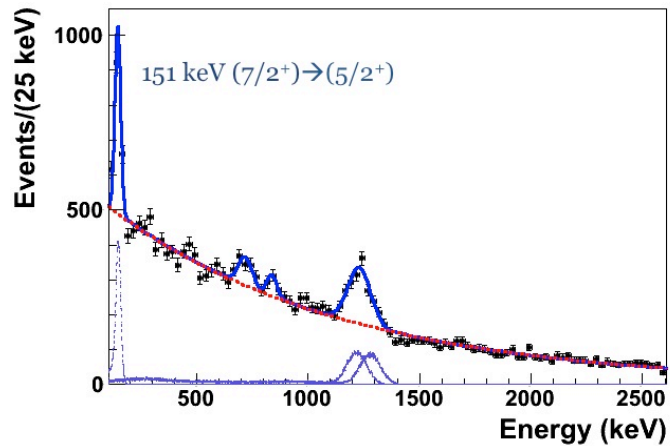

Figure 3: Doppler-corrected γ spectrum for the one-neutron knockout fragment $^{107}$Sn. The different widths of the peaks indicate doublets or multiplets as expected from the $^{107}$Sn spectrum around 1.2 MeV. In light blue the GEANT4 Monte Carlo simulation of the detector for the γ transition of interest.

Also for the $^{105}$Cd knockout residues a ML fit was performed to the γ spectra to construct level schemes through γ-γ coincidences. Figure 4 (top) shows the fit to the data up to 2 MeV for $^{105}$Cd. The energy level scheme of $^{105}$Cd is very complex due to multiple low energy gammas transitions, therefore, background subtraction and feeding need to be treated carefully. From studies of single and double coincidences, a preliminary energy level scheme was built, although limited by the energy resolution. A statistical technique called S-plot [17] was used to unfold the contributions of the different sources of background and isolate the events of interest. An event-by-event correction weight, defined by the covariant matrix of the fit to the data, was then applied to the momentum distribution. The momentum distributions obtained show $\ell=2$ and $\ell=4$ knockout by ground and first excited states, respectively, as would be expected for their 5/2+ and 7/2+ nature, as assigned in the literature [18].
This provides an important validation of the technique for the light tin analysis. Due to the intensity of the incoming $^{106}$Cd, two- and three-neutron knockout events were observed, as shown in Figure 4 (center and bottom). Preliminary inclusive cross sections of about 100 mb, 60 mb, and 20 mb were measured for one-, two-, and three-neutron knockout reactions, respectively.

## 3. Conclusions

Neutron knockout experiments performed at the NSCL on beams of $^{108,106}$Sn and $^{106}$Cd have improved our knowledge of the nuclear structure of nuclei around $^{100}$Sn. The spin-parity of the ground state of $^{107,105}$Sn have been found to be 5/2$^+$, while the one for first excited states to be 7/2$^+$. This result follows the systematics of heavier tin isotopes as expected from Shell Model calculation. The results also show a degree of mixing in the ground states of the isotopes $^{108,106}$Sn between the $d_{5/2}$ and $g_{7/2}$ single particle-states. For the $^{106}$Cd beam single-, double-, and triple-neutron knockout reactions have been observed. Since for $^{105}$Cd the spin-parity of the ground and first excited states is known, the studies performed confirm it and validate the technique used for the light tin isotopes. Preliminary inclusive cross sections have been measured for all the reactions and spectroscopic factors will be calculated from the momentum distributions.

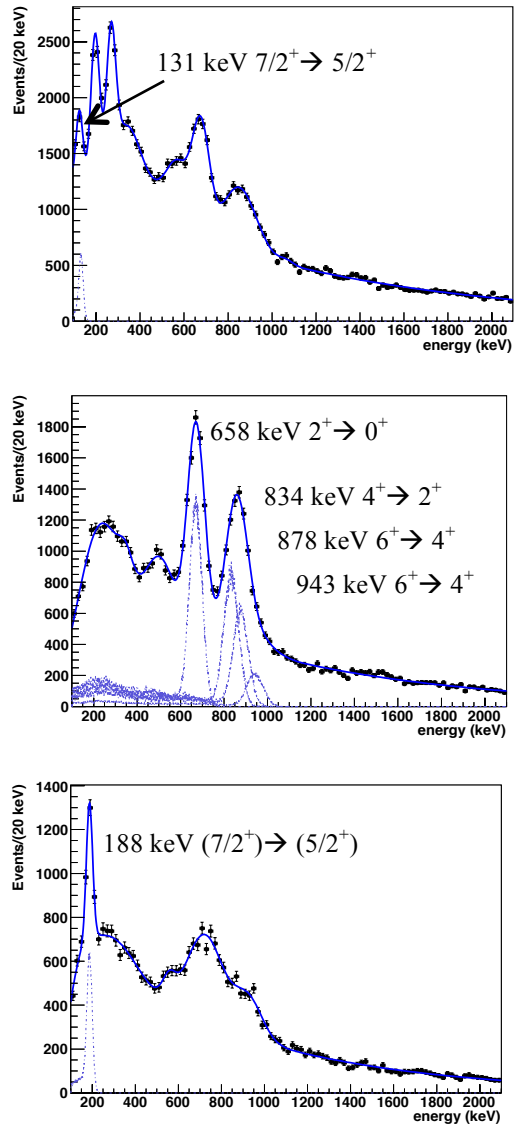

Figure 4: Doppler-corrected γ spectrum for the one- (top), two- (center), and three- (bottom) neutron knockout fragment $^{105,104,103}$Cd. The different resolutions

of the peaks indicate doublets or multiplets. In light blue the GEANT4 Monte Carlo simulation of the detector for the γ transition of interest.

## 4. Acknowledgement

This research was supported in part by the U.S. Department of Energy under grants DE-SC0001174 and DE-FE02-96ER-40983, the National Science Foundation under grants PHY-1067806 and PHY-1102511, and by the Basic Science Research Program through the National Research Foundation of Korea funded by the Ministry of Education, Science, and Technology under grant NRF-2012R1A1A1041763. I also want to thank the collaborators: A. Ayres, A. Bey, C. Bingham, L. Cartegni, G. Cerizza, K. Jones, R. Grzywacz, D. Miller, S. Padgett (UTK), T. Baugher, D. Bazin, J. Berryman, A. Gade, S. McDaniel, A. Ratkiewicz, A. Shore, R. Stroberg, D. Weisshaar, K. Wimmer, R. Winkler (MSU), A. Chae, S. Pain (ORNL), M.E. Howard (Rutgers), J. Tostevin (Surrey), and the A1900 and Cyclotron operators of the NSCL.